\documentclass[twoside]{article}
\usepackage{agupaper,amsmath}

\journalid{IJGA}{100}{10}{00.00.00} \cpright{IJGA}{2005}
\paperid{0}{1} \papercode{IJGA} \lefthead{Getling, Simitev, Busse}
\lefthead{Getling, Simitev, and Busse}\righthead{Global--Local
Solar Dynamo} \received{00.00.00} \revised{00.00.00}
\accepted{00.00.00} \online{00.00.00}{http://eos.wdcb.ru/ijga/}
\popref{yes} \keys{Sun}

\begin{document}
\title{Can cellular convection in a rotating spherical shell
maintain both global and local magnetic fields?}

\author{A. V. Getling}
\affil{Institute of Nuclear Physics, Lomonosov Moscow State
       University, 119992 Moscow, Russia}
\author{R. D. Simitev}
\affil{Department of Mathematical Sciences, The
  University of Liverpool, Liverpool L69~7ZL, UK}
\author{F. H. Busse}
\affil{Institute of Physics, University of Bayreuth, D-95440
       Bayreuth, Germany}

\abstract{A convection-driven MHD dynamo in a rotating spherical
shell, with clearly defined structural elements in the flow and
magnetic field, is simulated numerically. Such dynamos can be
called deterministic, in contrast to those explicitly dependent on
the assumed properties of turbulence. The cases most interesting
from the standpoint of studying the nature of stellar magnetism
demonstrate the following features. On a global scale, the
convective flows can maintain a ``general'' magnetic field with a
sign-alternating dipolar component. Local (in many cases, bipolar)
magnetic structures are associated with convection cells.
Disintegrating local structures change into background fields,
which drift toward the poles. From time to time, reversals of the
magnetic fields in the polar regions occur, as ``new'' background
fields expel the ``old'' fields.}

\section{Introduction}

The original motivation of this study was suggested by the
problems of solar physics. Although the results obtained at this
stage cannot be interpreted to represent the specifically solar
dynamo process, we should give a brief exposition of the previous
investigation that led us to the formulation of the problem
considered here.

Observations of the solar magnetic fields reveal a bewildering
variety of structures and activities. It is remarkable that solar
processes vary in scale from sizes comparable to the solar radius
to the limit of present resolution and in duration from tens of
years to minutes (see, for instance, \emph{Schrijver and Zwaan}
[2000]).

Mean-field electrodynamics [\emph{Moffatt}, 1978; \emph{Krause and
R\"adler}, 1980] has clarified many issues concerning the
generation of the global magnetic fields of cosmic bodies.
However, such problems as the formation of local magnetic fields
and their relationship to the global fields fall completely beyond
the scope of mean-field theories. Meanwhile, the phenomenon of
solar and stellar magnetism could be adequately understood only if
the dynamics of the interplay between structures in the velocity
field and magnetic field is comprehensively studied over a wide
range of spatial scales. Dynamo models that are aimed at a unified
description of the global and local processes and that deal with
local, instantaneous quantities rather than averaged ones can
naturally be referred to as ``deterministic'' models. They
describe the structural elements present in the flow and in the
magnetic field instead of considering the averaged parameters of
the turbulent flow (in particular, the statistical predominance of
one sign of the velocity-field helicity or another).

The idea that convection cells in the solar subphotospheric zone
could be a connecting link between global and local magnetic
fields traces back to the mid-1960s. \emph{Tverskoy} [1966]
represented the convection cell by a toroidal eddy and
demonstrated, in the framework of a kinematic approach, that such
a model convection cell can amplify the magnetic field and produce
characteristic bipolar magnetic configurations. This approach was
also used by \emph{Getling and Tverskoy} [1971] to construct a
kinematic model of the global dynamo in which toroidal eddies
distributed over a spherical shell, acting jointly with the
differential rotation of the shell, maintain a sign-alternating
global magnetic field. If a poloidal magnetic field is present,
the differential rotation produces a toroidal component of the
global magnetic field. If the local magnetic configuration
produced by an eddy interacting with the large-scale toroidal
field is rotated through some angle about the axis of the eddy,
this configuration contributes to the regeneration of the poloidal
component of the global magnetic field. Thus, a cell
\emph{locally} interacting with the magnetic field serves in this
model as a building block of the \emph{global} dynamo, and the
latter can in this case be called the ``cellular'' dynamo. The
rotation of the local magnetic-field pattern can be expected if
the system rotates as a whole and the flow is affected by the
Coriolis force.

In recent years, after the advent of suitable computing
facilities, some steps have been made to verify these ideas by
means of numerical simulation. \emph{Getling}~[2001] and
\emph{Getling~and Ovchinnikov}~[2002] obtained numerical solutions
to the three-dimensional nonlinear problem of magnetoconvection in
a plane horizontal layer of incompressible fluid, heated from
below, and found that hexagonal convection cells interacting with
a weak initial (``seed''), horizontal magnetic field can produce
various structures of the strongly amplified magnetic field, with
a predominant bipolar component. \emph{Dobler~and Getling}~[2004]
extended this numerical analysis to compressible fluids and
obtained similar results.

Modern computing resources make it possible to approach the
development of numerical cellular-dynamo models that could provide
a parallel description of both the global and local magnetic
fields. However, even today, numerical schemes can hardly be used
to simulate flows and magnetic fields over scale ranges covering
two or more orders of magnitude. Only the largest convection
cells~--- of sizes comparable to the depth of the convection zone
(such as solar ``giant'' cells, for which little observational
evidence exists)~--- can be simulated in the framework of global
models. If we assume that the principal features of the process
should be similar for convection on different scales, such global
models would help us to verify our qualitative notion and provide
guidelines for the elaboration of a more detailed description.

Here, we use numerical simulations to investigate the properties
of cellular dynamos in rotating spherical shells, which could
operate in stars under certain conditions. Although there are
reasons to believe that the solar dynamo is also of the cellular
type, we do not directly associate the presently obtained results
with solar processes, since the computed patterns of
magnetic-filed evolution bear only limited similarity to the
pattern observed on the Sun. We merely note some remarkable
features of the cellular dynamos, which may be of interest from
the standpoint of stellar magnetohydrodynamics.

\section{Formulation of the problem and numerical technique}

\begin{figure}
\captionwidth{\columnwidth}\centerline{
\includegraphics[width=6cm]{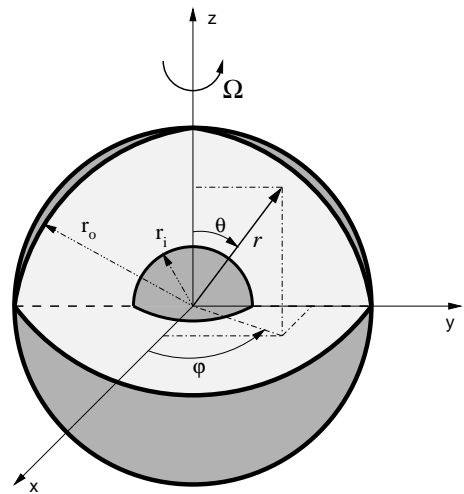}}
\caption{Geometrical configuration of the problem. The two
spherical surfaces with radii $r_\mathrm{i}$ and $r_\mathrm{o}$
are dark-shaded. A part of the outer surface is removed to expose
the interior of the shell (light-shaded) where the conducting
fluid is confined.} \label{f.01}
\end{figure}

In order to model the process of magnetic-field generation in a
stellar convection zone, we consider a spherical shell of
thickness $d = r_\mathrm{o} - r_\mathrm{i}$ (where $r_\mathrm{o}$
and $r_\mathrm{i}$ are the outer and inner radii of the shell),
full of electrically conducting fluid and rotating with a constant
angular velocity $\mathbf\Omega$ about a fixed axis $\hat {\mathbf
e}_z$, as shown in Figure~\ref{f.01}.

We follow the standard formulation used in earlier work by
\emph{Tilgner and Busse} [1997], \emph{Busse et al.} [1998],
\emph{Grote et al.} [1999, 2000], and \emph{Simitev and Busse}
[2002, 2005], but we assume a more general form of the static
temperature distribution,
\begin{align}
&T_\mathrm{S}=T_0 -\frac\beta 2 d^{\,2} r^2 + \frac{\beta_1}{d}
\frac 1 r, \label{profile}
\\ &\beta = \frac q{3\chi c_p}, \quad
\beta_1 = \frac{\eta d\Delta T}{(1-\eta)^2}, \nonumber
\end{align}
where the radial coordinate $r$ is measured in units of $d$,
$\chi$ is the thermal diffusivity, $c_p$ is the specific heat at
constant pressure, $q$ is the mass density of uniformly
distributed heat sources,  $\eta \equiv r_\mathrm{i}/r_\mathrm{o}$
is the inner-to-outer radius ratio of the shell, and $T_0$ is a
constant. The quantity $\Delta T$ is related to the difference
between the constant temperatures of the inner and outer spherical
boundaries, $T_\mathrm{i}$ and $T_\mathrm{o}$, as
\begin{equation}
\Delta T = T_\mathrm{i} - T_\mathrm{o} - \frac 1 2 \beta
d^{\,2}\frac{1+\eta}{1-\eta} \label{tempdiff}
\end{equation}
and reduces to $T_\mathrm{i} - T_\mathrm{o}$ in the case of $q=0$
(also dealt with in some simulations). The shell is
self-gravitating, and the gravitational acceleration averaged over
a spherical surface $r = \mathop{\mathrm{const}} \nolimits$ can be
written as $\mathbf g = - \gamma d \cdot \mathbf r$, where
$\mathbf r$ is the position vector with respect to the centre of
the sphere; as specified above, its length $r$ is measured in
units of $d$. In addition to $d$, the time $d^2 / \nu$, the
temperature $\nu^2 / \gamma \alpha d^4$ (where $\alpha$ is the
volumetric coefficient of thermal expansion), and the magnetic
induction $\nu ( \mu \rho )^{1/2} /d$ are used as scales for the
dimensionless description of the problem; here, $\nu$ denotes the
kinematic viscosity of the fluid, $\rho$ is its density, and $\mu$
is its magnetic permeability (we set $\mu=1$).

We use the Boussinesq approximation in that we assume $\rho$ to be
constant except in the gravity term, where, in addition to the
standard linear dependence $\rho(T)$  [according to which
$\rho^{-1}(d\rho/dT) = -\alpha = \mathop{\mathrm{const}}
\nolimits$], we introduce a small quadratic term in most cases.
Once a cellular pattern has developed, the presence of this term
and of the volumetric heat sources should not radically modify the
properties of the dynamo; however, both these factors favour the
development of polygonal convection cells [\emph{Busse,} 2004]
similar to the cells observed on the Sun, rather than meridionally
stretched, banana-like convection rolls. Without these essential
modifications, polygonal cells could only be obtained at much
smaller rotational velocities; in this case, the process would
develop very slowly, and the computations would be extremely
time-consuming.

Thus, the equations of motion for the velocity vector $\mathbf u$,
the heat equation for the deviation $\Theta$ from the static
temperature distribution, and the equation of induction for the
magnetic field $\mathbf B$ are

\begin{subequations}
\label{e.01}
\begin{align}
&\nabla \cdot \mathbf u = 0, \\
\label{e.01a} &(\partial_t  + \mathbf u \cdot \nabla) \mathbf u =
    - \nabla \pi
    + \tau \mathbf u \times \hat {\mathbf e}_z
    + (\Theta+ \epsilon\, \Theta^2) \mathbf r \nonumber \\
&\hspace{2.5cm} +  \nabla^2 \mathbf u
    + (\nabla \times \mathbf B) \times \mathbf B, \\
\label{e.01b} &P \left( \partial_t + \mathbf u \cdot \nabla\right)
\Theta  =
    \nabla^2 \Theta + \left(r_\mathrm{i} + R_e \frac{\eta}{(1-\eta)^2}
    \frac{1}{r^3}  \right)\mathbf r \cdot \mathbf u, \\
&\nabla \cdot \mathbf B = 0,  \\
\label{e.01c} &\partial_t \mathbf B =  \nabla \times
\left(\mathbf u \times
    \mathbf B \right) + P_m^{-1} \nabla^2 \mathbf B,
\end{align}
\end{subequations}
where $\pi$ is an effective pressure.

Six nondimensional \emph{physical} parameters of the problem
appear in our formulation. The Rayleigh numbers measure the energy
input into the system,
\begin{equation}
R_\mathrm{i}=\frac{\alpha \gamma \beta d^{\,6}}{\nu\chi}, \qquad
R_\mathrm{e}=\frac{\alpha \gamma \Delta T d^{\,4}}{\nu\chi},
\end{equation}
and are associated with the internally distributed heat sour\-ces
$q$ and the externally specified temperature difference
$T_\mathrm{i} - T_\mathrm{o}$ [see equation (\ref{tempdiff})],
respectively. The Coriolis number $\tau$, the Prandtl number $P$,
and the magnetic Prandtl number $P_\mathrm m$ describe ratios
between various time scales in the system,\
\begin{equation}
\tau=\frac{2\Omega d^{\,2}}{\nu}, \quad P=\frac{\nu}{\chi}, \quad
P_\mathrm m=\frac{\nu}{\nu_\mathrm{m}}
\end{equation}
($\nu_\mathrm{m}$ is the magnetic viscosity, or magnetic
diffusivity). Finally, $\epsilon$ is the small constant that
specifies the magnitude of the quadratic term in the temperature
dependence of density [see equation~(\ref{e.01a})].

\begin{figure}[t] \captionwidth{\columnwidth} \centering
\includegraphics[width=\columnwidth,bb=25 0 494 360pt,clip]{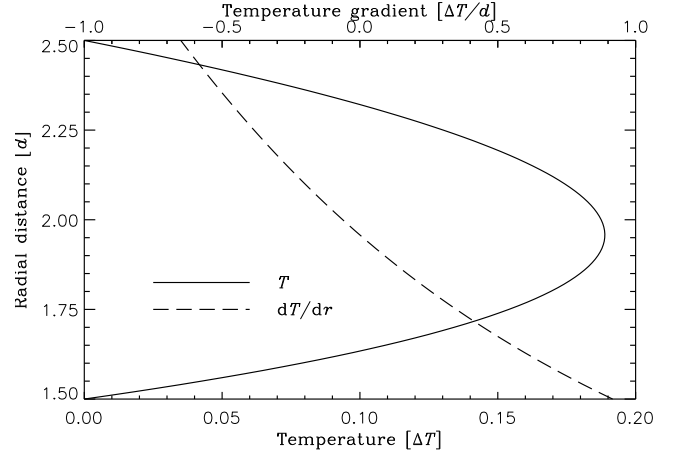}
 \caption{Static profiles of the temperature and temperature
gradient in the case of internal heating.} \label{profiles}
\end{figure}

Since the velocity field $\mathbf u$ and the magnetic induction
$\mathbf B$ are solenoidal vector fields, the general
representation in terms of poloidal and toroidal components can be
used,
\begin{subequations}
\begin{align}
&\mathbf u = \nabla \times ( \nabla v \times \mathbf r) + \nabla w
\times \mathbf r, \\ &\mathbf B = \nabla \times  ( \nabla h \times
\mathbf r) + \nabla g \times \mathbf r.
\end{align}
\end{subequations}
By multiplying the (curl)$^2$ and the curl of the Navier--Stokes
equation \eqref{e.01a} in the rotating system by $\mathbf r$, we
obtain two equations for $v$ and $w$,
\begin{subequations}
\label{e.02}
\begin{align}
&[( \nabla^2 - \partial_t) {\cal L}_2 + \tau \partial_{\varphi}]
\nabla^2 v + \tau {\cal Q} w - {\cal L}_2 (\Theta + \epsilon\,
\Theta^2) \nonumber \\ &\hspace{2.5cm}
 = - \mathbf r \cdot \nabla \times [ \nabla \times ( \mathbf u \cdot
\nabla \mathbf u - \mathbf B \cdot \nabla \mathbf B)] \\
&[( \nabla^2 - \partial_t) {\cal L}_2 + \tau \partial_{\varphi} ]
w - \tau {\cal Q}v = \mathbf r \cdot \nabla \times ( \mathbf u
\cdot \nabla \mathbf u - \mathbf B \cdot \nabla \mathbf B),
\end{align}
\end{subequations}
where $\varphi$ denotes the azimuthal angle (``longitude'') in the
spherical system of coordinates $r, \theta, \varphi$, and the
operators ${\cal L}_2$ and ${\cal Q}$ are defined by
\begin{displaymath}
{\cal L}_2 \equiv - r^2 \nabla^2 + \partial_r ( r^2 \partial_r) \\
\end{displaymath}
\begin{displaymath}
{\cal Q} \equiv r \cos \theta \nabla^2 - ({\cal L}_2 + r
\partial_r ) ( \cos \theta
\partial_r - r^{-1} \sin \theta \partial_{\theta}).
\end{displaymath}
The heat equation \eqref{e.01b} can be rewritten in the form
\begin{equation}
\label{e.03} \nabla^2 \Theta + \left[ r_\mathrm{i} +R_e \eta
r^{-3} (1 - \eta)^{-2} \right] {\cal L}_2 v = P ( \partial_t +
\mathbf u \cdot \nabla ) \Theta.
\end{equation}
Equations for $h$ and $g$ can be obtained multiplying the
equation of magnetic induction \eqref{e.01c} and its curl by
$\mathbf r$,
\begin{subequations}
\begin{align}
\label{e.04} &\nabla^2 {\cal L}_2 h = P_m [ \partial_t {\cal L}_2
h - \mathbf r \cdot \nabla \times ( \mathbf u \times \mathbf B )],
\\ &\nabla^2 {\cal L}_2 g = P_m [ \partial_t {\cal L}_2 g - \mathbf r
\cdot \nabla \times ( \ \times ( \mathbf u \times \mathbf B ))].
\end{align}
\end{subequations}

We assume stress-free boundaries with fixed temperatures,
\begin{equation}
\label{e.05}
v = \partial^2_{rr}v = \partial_r (w/r) = \Theta = 0 \mbox{ at }
 r=r_\mathrm{i} \mbox{ and } r=r_\mathrm{o}.
\end{equation}
For the magnetic field, we use electrically insulating boundaries
such that the poloidal function $h$ must be matched to the
function $h^{(\mathrm e)}$ that describes the potential fields
outside the fluid shell
\begin{equation}
\label{e.06} g = h-h^{(\mathrm e)} = \partial_r ( h-h^{(\mathrm
e)})=0 \mbox{ at } r=r_\mathrm{i} \mbox{ and } r=r_\mathrm{o} .
\end{equation}

\begin{figure}[p] 
\captionwidth{\columnwidth}\centering
\includegraphics[width=7cm]{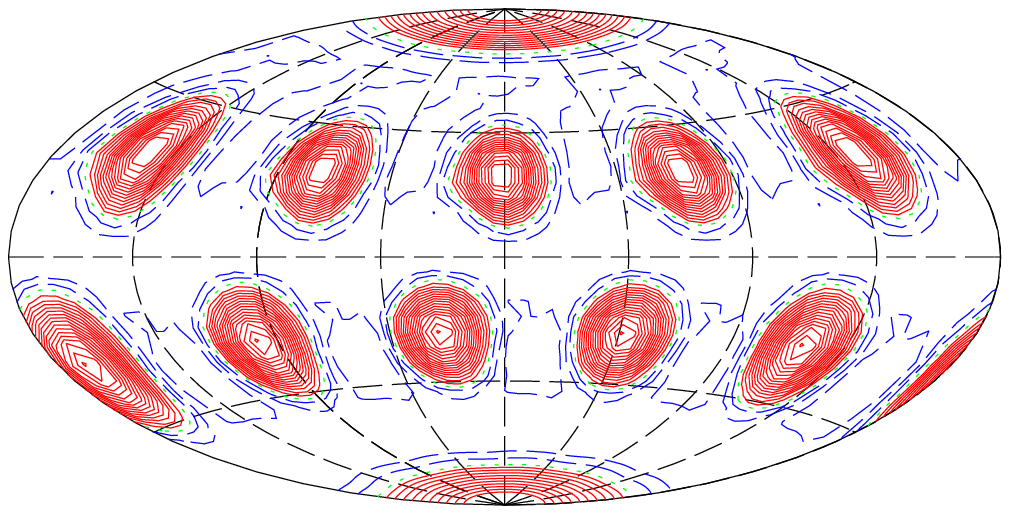}
\includegraphics[width=7cm]{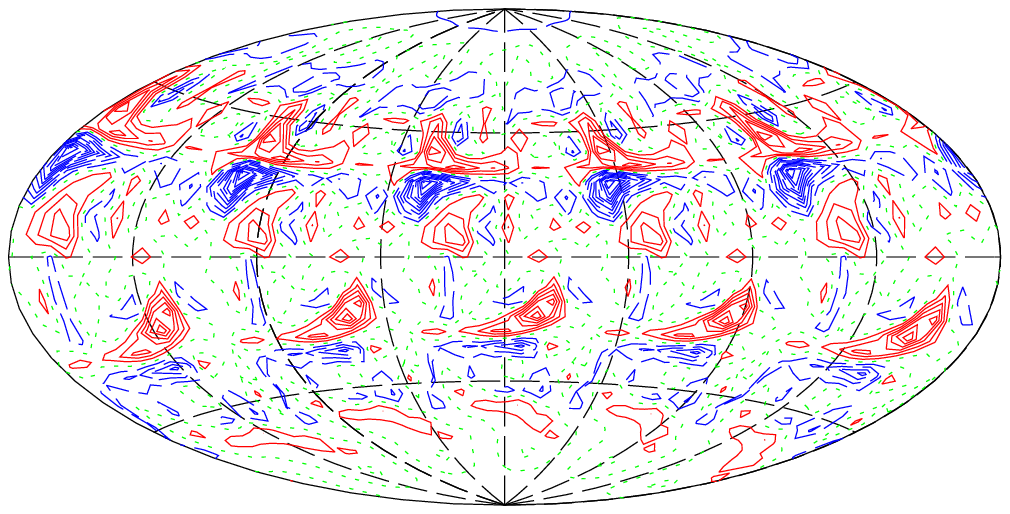}
\includegraphics[width=7cm]{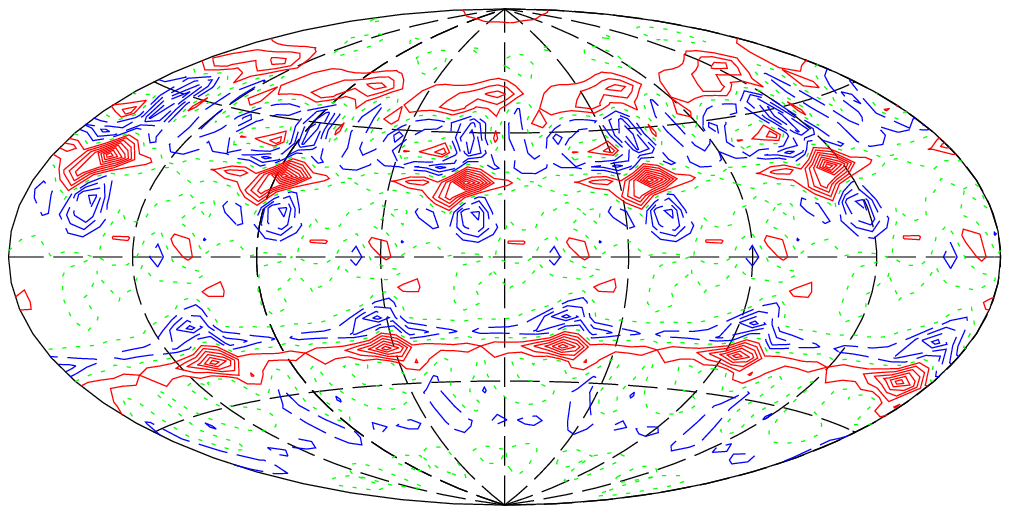}
\caption{Contours of the radial velocity component on the sphere
$r=r_\mathrm{i}+0.5$ at $t=98.73$ (top) and of the radial
component of the magnetic field on the sphere $r=r_\mathrm{o}$ at
$t=98.73$ (middle) and 101.73 (bottom) in the case of internal
heating at $\eta=0.6$, $R_\mathrm{i}=3000$, $R_\mathrm{e}=-6000$,
$\tau=10$, $P=1$, $P_\mathrm{m}=30$, and $m_0=5$. Solid (red)
curves: positive values; dotted (green) curves: zero values;
dashed curves (blue): negative values.} \label{radvelmagn}
\end{figure}

\begin{figure}[p] 
\captionwidth{\columnwidth}
\centering\includegraphics[width=5cm]{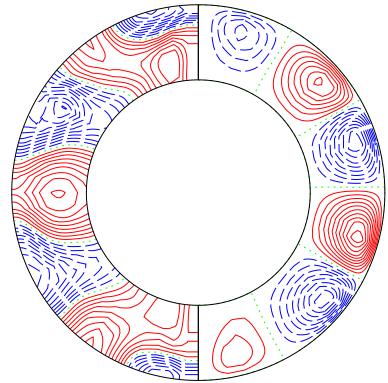}
\caption{Contours of the axisymmetric azimuthal velocity
(left-hand section) and streamlines of the meridional circulation,
or contours of the stream function of the axisymmetric meridional
flow (right-hand section) in the same case with internal heating
($P_\mathrm{m}=30$) at $t=98.73$. The curves have the same meaning
as in Figure~3.} \label{axisymm}
\end{figure}

The numerical integration of equations \eqref{e.02}--\eqref{e.06}
proceeds with a pseudospectral method developed by \emph{Tilgner
and Busse} [1997] and \emph{Tilgner} [1998], which is based on an
expansion of all dependent variables in spherical harmonics for
the $\theta$ and $\varphi$ dependences; in particular, for the
magnetic scalars,
\begin{subequations}\label{spher}
\begin{align}
&g=\frac 1r\sum_{l=0}^\infty \sum_{m=-l}^l G_l^m(r,t)P_l^m(\theta)
\exp \{ im \varphi \},
\\&h=\frac 1r\sum_{l=0}^\infty\sum_{m=-l}^l H_l^m(r,t)P_l^m(\theta)
\exp \{ im \varphi \}
\end{align}
\end{subequations}
(with truncating the series at an appropriate maximum~$l$), where
$P_l^m$ denotes the associated Legendre functions. For the $r$
dependences, truncated expansions in Chebyshev polynomials are
used. The equations are time-stepped by treating all nonlinear
terms explicitly with a second-order Adams--Bashforth scheme
whereas all linear terms are included in an implicit
Crank--Nicolson step.{\sloppy

}For the computations to be reported here, a minimum of 33
collocation points in the radial direction and spherical harmonics
up to the order 96 have been used. In addition to the
\emph{geometric} parameter $\eta$ and the above-mentioned
\emph{physical} parameters, we specified a \emph{computational}
parameter, viz., the fundamental (lowest nonzero) azimuthal number
$m_0$. Thus, only the following azimuthal harmonics were really
considered: $$1,\ \mathrm{e}^{\pm\mathrm{i} m_0\varphi},\
\mathrm{e}^{\pm 2 \mathrm{i} m_0\varphi},\ \mathrm{e}^{\pm 3
\mathrm{i} m_0\varphi}\ldots\,.$$ In other words, we imposed an
$m_0$-fold symmetry in the $\varphi$ direction. If $m_0\ne 1$,
this reduces the computation time.

\section{Results}

\subsection{Internal heating}

For the cases of internal heating, we varied $P_\mathrm{m}$ and
assumed $\eta=0.6$, $R_\mathrm{i}=3000$, $R_\mathrm{e}=-6000$,
$\tau=10$, $P=1$, and $m_0=5$. As can be seen from checking
computations with $m_0=1$ (not presented here), removing the
artificially imposed fivefold azimuthal symmetry does not
substantially modify the character of the convection pattern. The
quadratic term was present in the temperature dependence of
density, with a control parameter of $\epsilon=0.005$.

The distributions of the temperature $T_\mathrm{S}(r)$ and its
gradient $\mathrm{d}T_\mathrm{S}/\mathrm{d}r$ for the
corresponding static-equilibrium state are shown in
Figure~\ref{profiles}. Obviously, the outer part of the shell is
convectively unstable and the inner part is stable.

\subsubsection{The case of $\boldsymbol{P_\mathrm{m}=30}$.} At this
$P_\mathrm{m}$ value, the computations covered a time interval of
about 100 in units of the time of thermal diffusion across the
shell. Over most part of this period, a very stable pattern of
convection cells with a dodecahedral symmetry can be observed
(Figure~\ref{radvelmagn},\footnote{The colour figures can be
viewed in the online version of the paper.} top). These cells have
a normal appearance typical of cellular convection, without
substantial distortions due to the rotation of the shell. The
entire pattern drifts in the retrograde direction, in agreement
with theoretical predictions [\emph{Busse,} 2004].

\begin{figure} 
\captionwidth{\columnwidth}\centering
\includegraphics[width=\columnwidth]{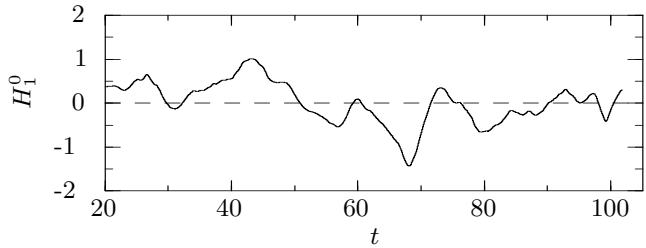}
\caption{Variation in the amplitude of the dipolar harmonic of the
poloidal magnetic field, $H_1^0(r,t)$, at $r=r_\mathrm{i}+0.5$ in
the same case with internal heating ($P_\mathrm{m}=30$).}
\label{dipole}
\end{figure}
\begin{figure*} 
\captionwidth{\textwidth}\centering
\includegraphics[width=\textwidth,height=4cm]{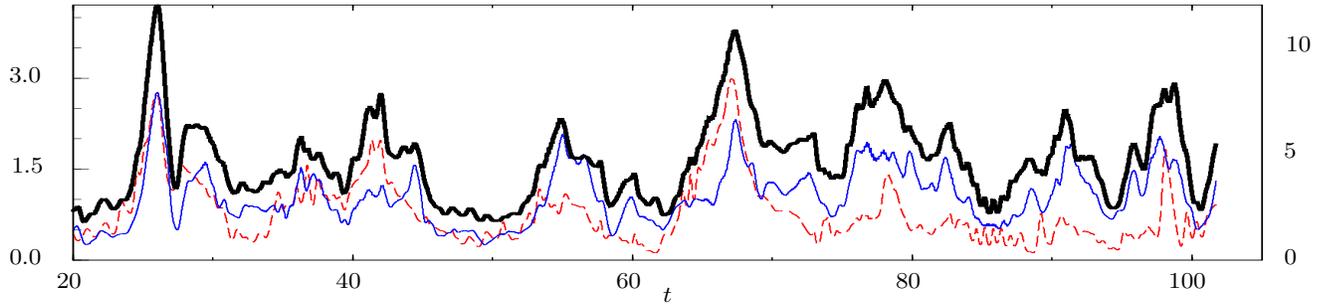}
\caption{Density of magnetic energy and its particular fractions
in the same case with internal heating ($P_\mathrm{m}=30$). Solid
(heavy black) curve and right-hand vertical scale: total density
of the magnetic energy; dashed (red) curve and left-hand vertical
scale: energy density associated with the axisymmetric part of the
magnetic-field component that has a dipole symmetry; dotted (blue)
curve and left-hand scale: energy density associated with the
nonaxisymmetric part of the same magnetic-field
component.}\label{energy}
\end{figure*}
The axisymmetric component of the azimuthal velocity
(Figure~\ref{axisymm}) in a well-established flow pattern is
nearly symmetric with respect to the equatorial plane.
Specifically, a prograde rotation of the equatorial zone (in the
frame of reference rotating together with the entire body) is
present along with a retrograde rotation of the midlatitudes, and
pairs of ``secondary'' prograde- and retrograde-rotation zones can
also be noted in the polar regions. In a nonrotating frame of
reference, the equatorial zone rotates more rapidly and the
midlatitudinal zones more slowly than the shell as a whole does.
Three pairs of meridional-circulation vortices fill the entire
meridional section of the shell, from one pole to another.

The pattern of magnetic field is less regular than the pattern of
flow (Figure~\ref{radvelmagn}, middle and bottom). Some remarkable
features or the simulated dynamo process can be summarized as
follows.

\begin{figure*}[t] 
\captionwidth{\textwidth}\centerline{
\includegraphics[width=7cm]{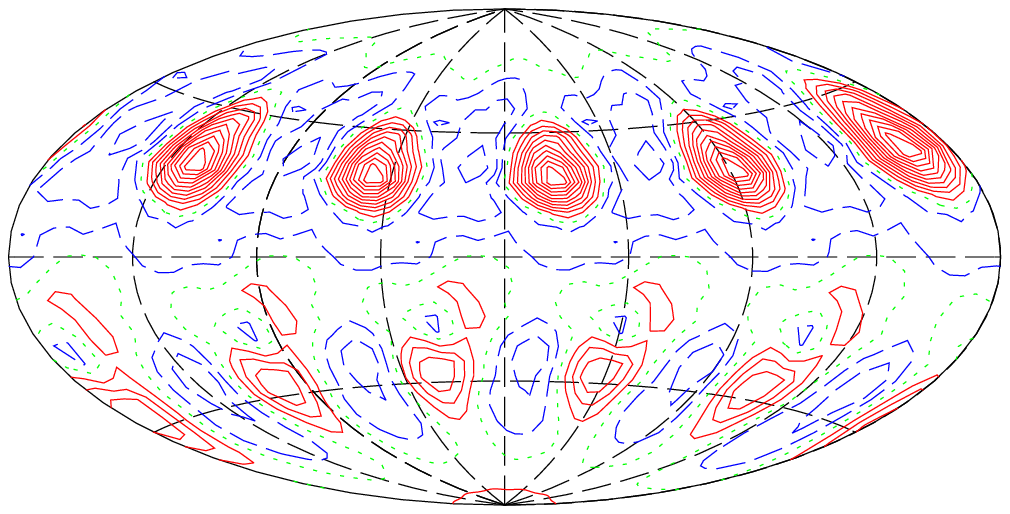}\quad\quad\quad
\includegraphics[width=7cm]{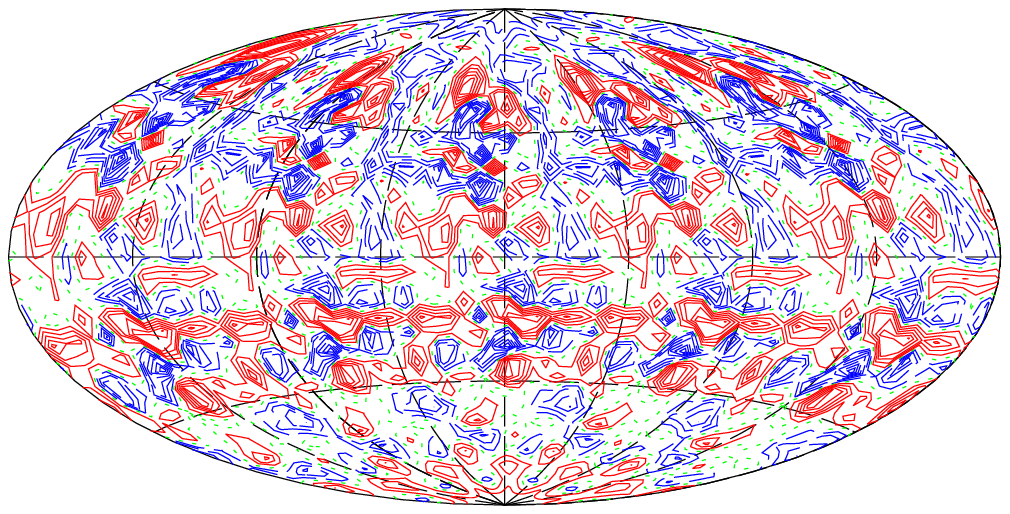}}
\centerline{
\includegraphics[width=7cm]{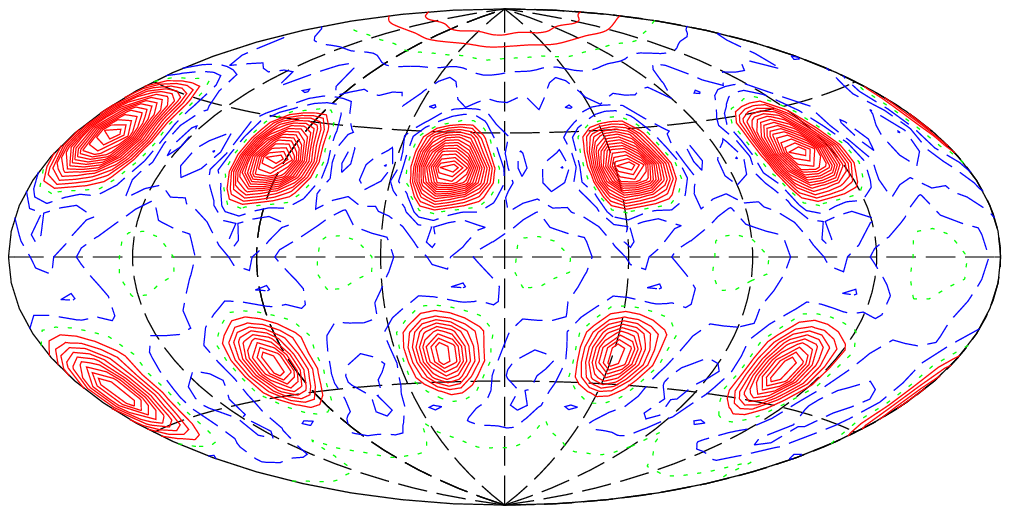}\quad\quad\quad
\includegraphics[width=7cm]{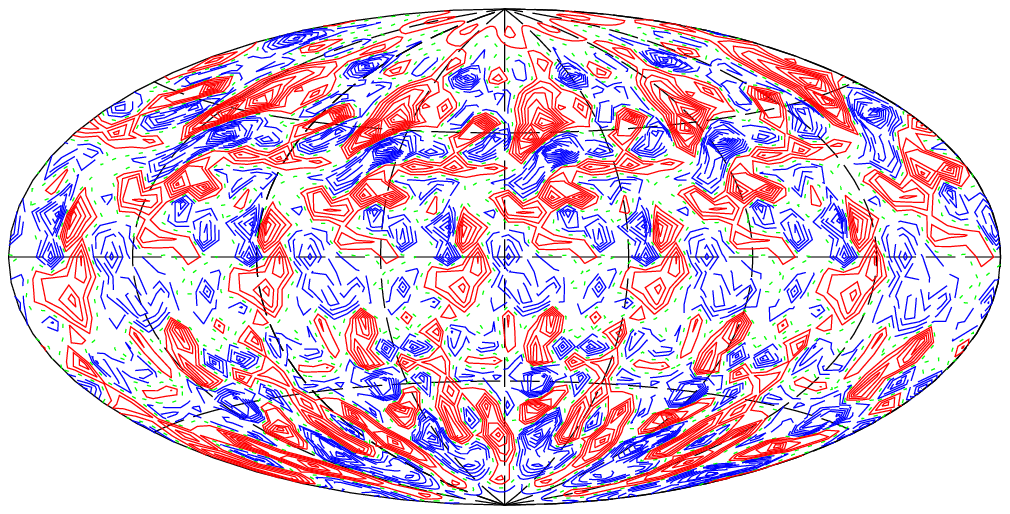}}
\caption{Contours of the radial component of the velocity on the
sphere $r=r_\mathrm{i}+0.5$ (left) and of the magnetic field on
the sphere $r=r_\mathrm{o}$ (right) at $t=200$ (top) and $t=327.2$
(bottom) in the case of internal heating at $P_\textrm{m}=200$,
while the other parameters are as in the preceding case:
$\eta=0.6$, $R_\mathrm{i}=3000$, $R_\mathrm{e}=-6000$, $\tau=10$,
$P=1$, and $m_0=5$. The curves have the same meaning as in
Figure~3.} \label{radmagn_p200}
\end{figure*}

\begin{figure*} 
\captionwidth{\textwidth}\centering
\includegraphics[width=\textwidth,height=4cm]{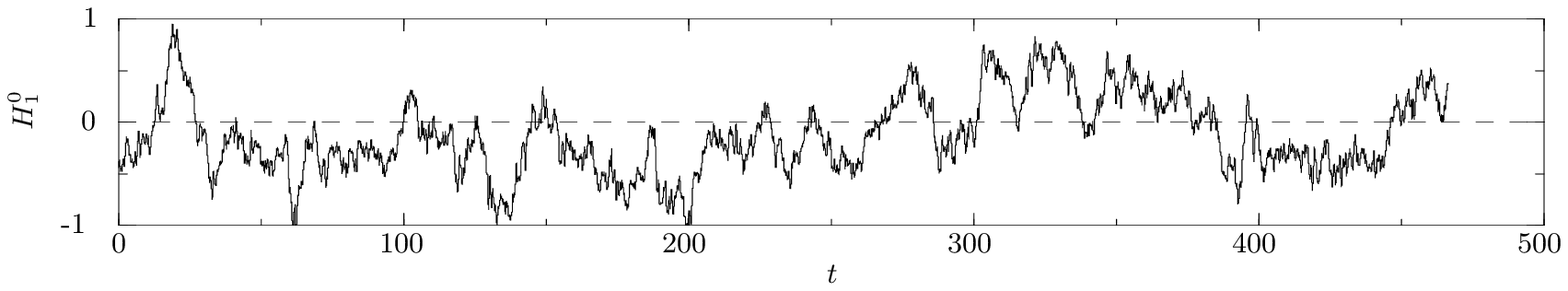}
\caption{Variation in amplitude of the dipolar harmonic of the
poloidal magnetic field, $H_1^0(r,t)$, at $r=r_\mathrm i +0.5$ for
the same case with internal heating as illustrated by
Figure~\ref{radmagn_p200} ($P_\mathrm{m}=200$).}
\label{dipole_p200}
\end{figure*}

First, local magnetic structures associated with convection cells
emerge repeatedly as compact magnetic regions (see
Fig.~\ref{radvelmagn}). In their subsequent evolution, these
regions change their configuration and finally dissipate into much
weaker remnant fields.

Second, the dipolar component of the global magnetic field
exhibits polarity reversals [see Figure~\ref{dipole} for a graph
of the amplitude of the dipole component,
$H_1^0(t)$]\footnote{Wherever the $r$ variable as an argument of
$H_1^0$ is omitted, we mean $H_1^0(r_\mathrm i+0.5,t)$}. The
background fields~--- remnants of the decaying local magnetic
structures~--- drift toward the poles and ``expel'' the ``old''
background fields present in the polar regions. As a result, the
old magnetic polarity is replaced with the new one due to the
poleward drift of the latter. The polarity reversals of the global
magnetic field can also be seen from the variation in the
amplitude of the dipolar harmonic of the poloidal field,
$H_1^0(t)$ (Figure~\ref{dipole}). The two maps of the magnetic
field shown in Figure~\ref{radvelmagn} correspond to two
situations in which the global magnetic dipole has opposite
orientations (the polarity reversal between these two times
corresponds to the rightmost intersection of the curve in
Figure~\ref{dipole} with the horizontal zero line).

Third, an interesting intermittent behaviour is exhibited by the
magnetic energy of the system. Let us compare the full energy and
two particular fractions of the energy associated with the
magnetic-field component that has a dipolar-type symmetry (i.e.,
is antisymmetric with respect to the equatorial plane).
Specifically, we are interested in the behaviour of the energy of
the axisymmetric and the nonaxisymmetric part of this component.
The axisymmetric part is represented by the spherical harmonics
with $l$ odd and $m=0$ [see (\ref{spher})], and the
nonaxisymmetric part by other harmonics with $l+m$ odd. As can be
seen from Figure~\ref{energy} (in which the total energy and its
particular fractions are divided by the volume of the shell), the
main peaks in the graph of the total energy are alternately
associated with increases in the energies of the axisymmetric and
the nonaxisymmetric part of the component with a dipolar symmetry.
In particular, the peak located near $t=42$ is fed by the
symmetric field; near $t=55$, by the asymmetric field; near
$t=68$, by both but with some predominance of the symmetric part;
and near $t=78$, again by the asymmetric part.

\subsubsection{The same case of $\boldsymbol{P_\mathrm{m}=30}$
but with special initial conditions.} In our attempts to find
conditions for the realization of magnetic-field dynamics similar
to the generally imagined pattern of a hypothetical dynamo process
with differential rotation as its essential part (known since the
qualitative model suggested by \emph{Babcock} [1961] and
\emph{Leighton} [1964, 1969]), we made an additional computational
run. We specified all parameters to be the same as in the
above-described case. However, the initial conditions were chosen
in such a way that, initially, the system would more likely find
itself within the attraction basin of the expected dynamo regime
in state space. To this end, we superposed the nonaxisymmetric
components of the velocity field and magnetic field obtained in
the above-described run onto a pattern of differential rotation
with the equatorial belt accelerated compared to higher latitudes
and with an appropriate distribution of the axisymmetric azimuthal
magnetic field that has different signs on the two sides of the
equator.{\sloppy

}The computations have demonstrated that the system nevertheless
approaches virtually the same regime that was observed without
using such special initial conditions. Thus, a closer similarity
between the numerical solution and the properties of the
hypothetical dynamos of the Babcock--Leighton type does not seem
to be achievable in the framework of this very simple model.

\subsubsection{The case of $\boldsymbol{P_\mathrm{m}=200}$.}
As the magnetic Prandtl number $P_\mathrm m$ is varied (under
otherwise fixed conditions), the convection pattern varies little
over a fairly wide $P_\mathrm m$ range. However, the greater this
parameter, the higher the mean strength of the magnetic field
(and, accordingly, the total magnetic energy). The kinetic energy
of convection decreases with the increase of $P_\mathrm m$ and
convection becomes more sensitive to time variations in the
magnetic field. The increase of $P_\mathrm m$ is also manifest in
the fact that local magnetic fields become more patchy and less
ordered. Individual areas filled with the magnetic field of a
given sign are smaller in size and more numerous, and bipolar
structures are not so well pronounced (see
Figure~\ref{radmagn_p200}, which refers to $P_\mathrm{m}=200$). As
in the case of $P_\mathrm m=30$, we can observe the penetration of
background fields into the polar regions and sign reversals of the
polar background fields.

Our computation for $P_\mathrm m=200$ cover a time interval almost
five times as long as for $P_\mathrm m=30$
(Figure~\ref{dipole_p200}). The two velocity maps and two
magnetic-field maps shown in Figure~\ref{radmagn_p200} nearly
correspond to the times of one negative and one positive extremum
of the amplitude $H_1^0(t)$ (see Figure~\ref{dipole_p200}). It is
remarkable that the polar background fields have different
polarities at these two times. At $t=200.0$, the background
magnetic field is negative in the ``northern'' and positive in the
``southern'' polar region; an opposite situation takes place at
$t=327.2$. The $H_1^0(t)$ curve demonstrates numerous sign
reversals, although fine details of this dependence only reflect
the irregular, fluctuational aspect of the process. It is
nevertheless clear that, even if we smooth this curve, it will
exhibit quite pronounced cyclic, although nonperiodic, polarity
reversals of the dipolar component of the ``general'' magnetic
field.

It should be noted that the distribution of the axisymmetric
component of the azimuthal velocity  (the pattern of differential
rotation, not shown here) in this case is much more complex and
variable than at $P_\mathrm m=30$. This effect also can be due to
the stronger influence of the magnetic field on the fluid motion.

\subsubsection{Internal heating without a $\boldsymbol{\Theta^2}$
term (nonmagnetic case).} To form an idea of the role played by
the quadratic term in the $\rho(T)$ dependence, we computed a
purely hydro\-dynamic (with $\mathbf B=0$) scenario under
conditions that differed from the conditions of the
above-described simulation by the absence of the $\Theta^2$ term
($\epsilon=0$) and by the Coriolis number ($\tau=1$); in addition,
we assumed $m_0=1$ in this case.

\begin{figure}[p]
\captionwidth{\columnwidth} \centering
\includegraphics[width=7cm,height=3.5cm]{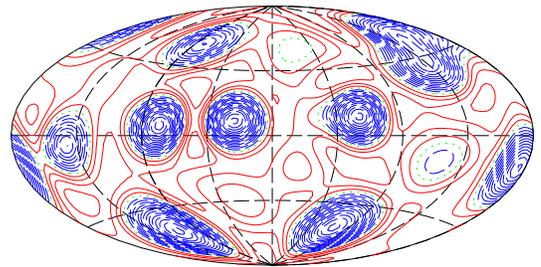}
 \caption{Contours of radial velocity on the surface
 $r=r_\mathrm i+0.5$   
  in the nonmagnetic case with internal heating, $\tau = 1$; no quadratic
  term is present in the $\rho(T)$ dependence ($\epsilon = 0$). The
  other parameters are $\eta=0.6$, $R_\mathrm{i}=3000$,
  $R_\mathrm{e}=-6000$, $P=1$ (as in the preceding cases),
  and $m_0=1$. The curves have the same meaning as in Figure~3.}
\label{nonmagn}
\end{figure}

The principal result of these computations is the finding that, in
the absence of the quadratic term, convection unaffected by the
magnetic field forms patterns of well-localized, three-dimensional
cells, which typically appear as shown in Figure~\ref{nonmagn}. A
downwelling is observed in the centre of each cell, in contrast to
the above-described cases, in which central upwellings typically
developed. The issue of the direction of circulation in a
convection cell is a fairly subtle matter (see, e.g.,
\emph{Getling} [1999] for a survey of some situations related to
convection in horizontal layers), so that agreement or
disagreement between our model and any really observed pattern can
in no way be an indication for the factors responsible for the
observed direction of convective motions. Our primary interest in
the cases where an upwelling is present in the central part of a
cell is merely dictated by our intention of constructing a dynamo
model reproducing the solar phenomena as closely as possible.

\begin{figure}[p]
\captionwidth{\columnwidth}\centering
\includegraphics[width=7cm,height=3.5cm,bb=154 468 440 612,clip]
{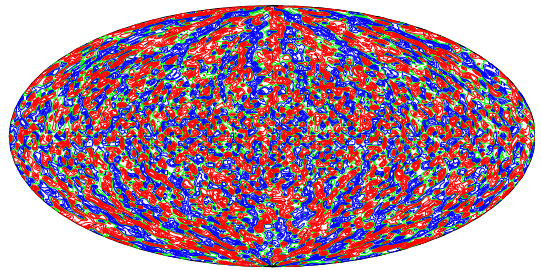} \caption{Map of the radial velocity component
on the sphere
$r=r_\mathrm{i}+0.5$ 
in the nonmagnetic case with
heating through the inner boundary at $\eta=0.9$, $P=1$,
$\tau=0.1$, $R_\mathrm e = 5000$, and $m_0=2$. The contours merge
into hatching: moderately light (red) for positive values, light
(green) for zero values, and dark (blue) for negative values.}
\label{radvel_t1_q=0}
\end{figure}

\subsection{Heating ``from outside'' (through the inner surface)}

In addition, we undertook a search for regimes in which convection
preserves its ``three-dimensional'' structure in the absence of
internal heat sources. In other words, some computations were done
at $R_\mathrm i=0$. The quadratic term was also missing from the
temperature dependence of density ($\epsilon = 0$) in these runs.

Note that, in the limiting case of a nonrotating shell, convection
cells are not stretched in any direction. Therefore, a cellular
pattern of convective motion can obviously be maintained even
without such favourable factors as a specific form of
stratification and a quadratic term, but at smaller rotational
velocities.

In shells without internal heat sources ($q=0$), convection
regimes similar to those observed at $q\ne 0$ should be expected,
under otherwise identical conditions, at smaller $\eta$. This is
because a stratification similar to that shown in
Figure~\ref{profiles} confines the development of convective
motion to the outer part of the shell.

We illustrate here the case of $q=0$ only by a tentative
computational run for convection without a magnetic field, at
$\eta=0.9$, $P=1$, $\tau=0.1$, $R_\mathrm e = 5000$, and $m_0=2$
(see Figure~\ref{radvel_t1_q=0} for a velocity field typical of
this case). Although the convection pattern is complex in this
case, a tendency toward the formation of meridionally elongated
cells can nevertheless be noted.

Computations with the magnetic field included (e.g., for
$\eta=0.8$, $P=1$, $R_\mathrm e=5000$, $\tau=0.1$, $P_\mathrm
m=5$) demonstrate the development of magnetic features with a very
small spatial scale, close to the resolution limit of the
computational scheme. This is a signature for an insufficient
spatial resolution, so that the simulation results are not quite
reliable.

The qualitative aspects of the results suggest that the Coriolis
number proves again to be insufficiently small for the stability
of ``three-dimensional'' convection cells, and the cells
ultimately become substantially stretched, although not in a
strictly meridional direction.

On the whole, the last two scenarios of flow and magnetic-field
evolution indicate that regimes of ``cellular'' dynamo in a shell
without internal heating and without a quadratic term in the
$\rho(T)$ dependence should be sought in the range of smaller
$\Omega$ (and $\tau$).

\section{Conclusion}

We have constructed relatively simple numerical models that
describe a self-sustained process of generation of interacting
global and local magnetic fields. As in most hypothetical
astrophysical dynamos, the generation is driven by thermal
convection in combination with differential rotation. We did not
introduce any kinematic elements in our model, so that the entire
velocity field appeared as the solution of the full system of MHD
equations.

The most remarkable features revealed in the computed dynamo
regimes can be summarized as follows. The process of
magnetic-field generation is cyclic, although rather irregular. It
includes the repeated generation of local, in many cases bipolar,
magnetic structures. These structures dissipate giving rise to
chaotic background fields. They may drift in the poleward
direction, replacing the already existing, ``old'' background
fields. In some cases, a correspondence can be noted between such
polarity reversals in the polar regions and sign reversals of the
axisymmetric bipolar component of the global magnetic field.

One of the computed scenarios demonstrates a remarkable
intermittency in the behaviour of some fractions of magnetic
energy: the axisymmetric and the nonaxisymmetric part of the
magnetic-field component with a dipolar symmetry alternate in
making larger contributions to the total-energy peaks.

Mean-field dynamo models, which have been most popular in
astrophysics over a few past decades, attribute the generation of
the global magnetic fields of stars to the $\alpha$ effect -- the
statistical predominance of one sign of the velocity-field
helicity over another. It is quite plausible that the $\alpha$
effect, in one form or another, is a fairly general property of
various velocity fields capable of maintaining undamped regular
magnetic fields. However, this property must not necessarily be
associated with turbulent motion. In particular, the model
velocity field in the toroidal eddies used by \emph{Getling and
Tverskoy} [1971] to construct a global dynamo included an
azimuthal (with respect to the axis of the eddy) velocity
component, so that the trajectories of the fluid particles were
spirals deformed in a certain way. A similar property may also be
inherent in the convective flow that develops in our model,
although checking this possibility requires a special
investigation.

At this stage, the ``deterministic'' cellular dynamo described
here is oversimplified to be regarded as a model of the solar or
any other specific stellar dynamo. However, it demonstrates that
the dynamics of the well-structured local magnetic fields and of
the global magnetic field may be ingredients of one complex
process.

\acknowledgements{The work of A.V.G. was supported by the Deutscher
Akademischer Austauschdienst, European Graduate College
``Non-Equilibrium Phenomena and Phase Transitions in Complex
Systems,'' and Russian Foundation for Basic Research (project code
\mbox{04-02-16580}).}

\end{document}